\documentclass[showpacs,twocolumn,prb,aps]{revtex4}
\usepackage{epsfig}
\usepackage{amsmath}
\usepackage{amssymb}
\begin{document}
\title{Diverging magnetothermal response in the one-dimensional
       Heisenberg chain}

\author{Kim Louis and C. Gros} 

\affiliation{Fakult\"at 7, Theoretische Physik,
 University of the Saarland,
66041 Saarbr\"ucken, Germany.}

\date{\today}

\begin{abstract}
A current of magnetic moments will flow in the
spin-1/2 Heisenberg chain in the presence of an
external magnetic field $B$ and a temperature gradient
$\Delta T$ along the chain. We show that this 
magnetothermal effect is strictly {\em infinite}
for the integrable Heisenberg-model in one dimension.
We set-up the response formalism and derive several
new generalized Einstein relations for this
magnetothermal effect which vanishes in the absence
of an external magnetic field. We estimate the size
of the magnetothermal response by exact diagonalization
and Quantum Monte Carlo and make contact with recent
transport measurements for the one-dimensional 
Heisenberg compound $\rm Sr_2CuO_3$.

\end{abstract}
\pacs{75.30.Gw,75.10.Jm,78.30.-j}
\maketitle


\section{Introduction} \label{sec_intro}

The nature of magnetic and thermal transport in 
magnetic insulators with
reduced dimensions is a long standing problem.
Huber, in one of the first works on the subject, \cite{hub68}
evaluated the thermal conductivity $\kappa(T)$ for the
Heisenberg-chain with an equation-of-motion approximation
and found a {\em finite} $\kappa(T)$, a result which is,
by now, known to be wrong. 
A few years later,  Niemeijer and
van Vianen calculated $\kappa(T)$ in the case $J_z=0$ 
using  the Jordan-Wigner transform and found a diverging result.\cite{NieVia71}
It has been shown
recently, \cite{ZNP} that the energy-current operator
commutes with the Hamiltonian for the
spin-1/2 Heisenberg chain. The thermal conductivity
is consequently {\em infinite} for this model.
The intriguing question, ``under which circumstances
does an interacting quantum-system show an
infinite thermal conductivity'', is 
being intensively studied theoretically, \cite{AlGr_kappa,hei02,KlSa}
motivated, in part, by new experimental findings.

An anomalous large
magnetic contribution to $\kappa$ has been observed for the spin-ladder system 
Sr$_{14-x}$Ca$_x$Cu$_{24}$O$_{41}$ and
Ca$_{9}$La$_{5}$Cu$_{24}$O$_{41}$, \cite{sol00,hes01} raising
the possibility of ballistic magnetic transport limited
only by residual spin-phonon and impurity scattering.
Large energy-relaxation times have been found in
recent experimental \cite{Sol01} and theoretical
\cite{AlGr_kappa} studies of the
thermal conductivity for the quasi-one dimensional
spin-chain compounds $\rm SrCuO_2$ and $\rm Sr_2CuO_3$,
consistent with $^{17}O$ NMR studies. \cite{thu01}

It is known, \cite{ZNP,ChLu} that there is no
magnetothermal effect in the Heisenberg chain since
the magnetic-current operator is a pseudo-vector.
This can be seen by a simple symmetry argument
in the standard setting, when we consider only a temperature 
gradient along the sample. A non-zero magnetothermal effect 
would yield a magnetization along an arbitrary quantization axis.
But because of the isotropic conditions there is no preferred direction
along which such a magnetization might occur, the effect
vanishes consequently.

\begin{figure}
\centerline{
\epsfig{file=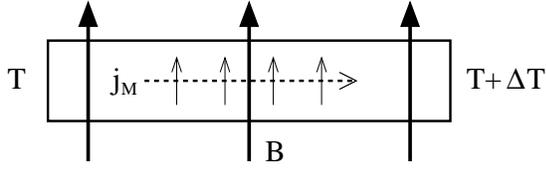,width=0.4\textwidth,angle=0}
           }
\caption{Illustration of a quasi-one-dimensional magnetic insulator
         in the presence of an external magnetic field $B$, and a
	 longitudinal temperature differential $\Delta T$.
	 A magnetic current $j_M$, with the moments aligned 
	 along $B$ is induced. }
\label{fig_illu}
\end{figure}

The situation is, however, different if there is an 
external magnetic field \cite{ZNP}. A temperature gradient 
will now cause a magnetization current with a magnetization
vector parallel to the field, see Fig.\ \ref{fig_illu}.  
This magnetothermal effect
can be considered the generalization to magnetic systems
of the Seebeck effect occurring in normal conductors.
It is diverging for the
isolated Heisenberg chain, leading to a finite non-zero
`magnetothermal Drude weight'. 
In this paper we discuss magnetothermal effects in spin chains,
i.e., we examine how the magnetic and thermal currents couple 
to external sources. 

In Sec.\ \ref{sec_response} we will discuss the
differences in between the thermomagnetic effects
found in normal metals and the magnetothermal response
we study here for magnetic insulators.
In Sec.\ \ref{sec_operators}, \ref{sec_corr_func}
and in Sec.\ \ref{sec_coeff}
we set-up the formalism for the magnetic and energy
current operators in the presence of an external 
magnetic field and for the correlation and response-functions,
respectively.

In Sec.\ \ref{sec_jordan} we evaluate the magnetothermal 
response for the xy-chain via a Jordan-Wigner transformation
and derive and discuss several generalized Einstein relations
in Sec.\ \ref{sec_einstein}. In Sec.\ \ref{sec_exact_Einstein}
we derive and discuss a new exact magnetothermal 
Einstein relation.

 We present
in Sec.\ \ref{sec_num_res} 
numerical results obtained from Quantum-Monte-Carlo Simulations
and exact-diagonalization studies.
Finally, we discuss the size of the magnetothermal effect
expected for $\rm Sr_2CuO_3$ together with a dimensional analysis
in Sec.\ \ref{sec_size}.

\section{Thermomagnetic and magnetothermal response} 
\label{sec_response}

In a conventional  solid there is normally a variety 
of ways in which external parameters (e.g., current density $j$,
temperature gradient and magnetic field $B$) couple to 
the voltage $\nabla\mu$ and the thermal current $j_E$
leading to numerous thermoelectric and thermomagnetic effects.
A phenomenological equation which gives full account of all these
effects would look as (see, e.g., Ref.\ \onlinecite{Abrikosov})

\begin{displaymath}
\left(\begin{array}{c}-\mbox{\boldmath$\nabla$}\mu\\ 
{\bf j}_E\end{array}\right)=
\left[
\left(\begin{array}{cc}\rho & Q\\ 
\Pi&-\kappa\end{array}\right)
+\left(\begin{array}{cc}R&N\\ 
E\kappa& L\kappa\end{array}\right)
{\bf B}\times\right]
\left(\begin{array}{c} {\bf j}\\
 \mbox{\boldmath$\nabla$} T\end{array}\right).
\end{displaymath}
(${\bf B}\times$ is understood to be applied to both components separately.)
$\rho$ is the resistivity, $\kappa$ the thermal conductivity,
and the other coefficients describe the
Seebeck- ($Q$), Peltier- ($\Pi$), Hall- ($R$), Nernst- ($N$), Ettingshausen-
($E$), and Righi-Leduc-effect ($L$). The latter four effects
are called thermomagnetic -- for obvious reason -- 
and are not to be confounded with the effects which are to be discussed
in this paper which we denote {\em magnetothermal effects}.
One should note that in magnetic systems
the magnetic field corresponds to the chemical
potential, leaving  no analogues for the entries of the second matrix.

Following Ref.\ \onlinecite{Abrikosov}, we write the dependence of the
current densities  $j_M$ (particle) and $j_E$ (energy)  on external sources
(field $\nabla B=-\nabla\mu$ and temperature gradient $\nabla T$)
in the following manner

\begin{equation}\label{Ltensor}
\left(\begin{array}{c}j_M\\ j_E\end{array}\right)=
\left(\begin{array}{cc}\hat L_{\mathit MM}&\hat L_{\mathit ME}\\ 
\hat L_{\mathit EM}&\hat L_{\mathit EE}\end{array}\right)
\left(\begin{array}{c}\nabla B\\
(-\nabla T)\end{array}\right).\end{equation}

[We deviate from the standard notation by indexing the components
of the tensor by $M$ and $E$ rather than 1 and 2. The notation with $M$ as
particle index and $B$ as chemical potential borrows from spin systems,
this article's chief case of interest. Eq.\ (\ref{Ltensor}) may therefore 
be used as a
reference for both electrical and magnetic systems.]
Normally, the two off-diagonal components governing the {\em Seebeck} 
and {\em Peltier} effect are not independent but related via the
{\it Onsager relation}.  Here, with the  given choice of external forces, it
reads (where $T$ denotes temperature)
\begin{equation}
T\hat L_{\mathit ME}\ =\ \hat L_{\mathit EM}~.
\label{onsager}
\end{equation}

To compute the response functions
$\hat L_{ij}$ we use the standard Kubo-formula
\cite{NieVia71,KlSa,AlGr_MC,ZMR} which yields
for the Heisenberg model
generally an $\hat L$-Tensor with infinite components,
i.e., a delta function times a weight-factor.
These weight-factors will be denoted by 
the entries of the $L$-Tensor 
$L_{ij}$ (without $\hat{\phantom a}$-accent).
Normally, one has an additional finite contribution -- the regular
part -- which will turn out to be zero for all but the MM-component.
We assume that due to some -- so far unaccounted -- 
scattering processes with external degrees of freedoms
(like phonons or impurities)
the infinite peak broadens, and the
coefficients (we consider only the real part) may be replaced according
to (cf., e.g., Ref. \onlinecite{AlGr_MC})
\begin{equation}
\hat L_{ij}(\omega)\ \equiv\ L_{ij}\pi\delta(\omega)\to
\frac{L_{ij}\tau}{1+(\omega\tau)^2}
\label{L_tau}
\end{equation}
with $i,j\in\{M,E\}$ and a finite relaxation time
$\tau$ -- for simplicity,  a possible dependence on $i$ and $j$ shall
 be neglected.

\section{Current operators} \label{sec_operators}

In this article, we discuss the analogue of the thermoelectric
effect in spin-chains.
The Hamiltonian is the standard $xxz$-chain, $H=\sum_n h_n$ with

\begin{equation}
h_n\ =\ J_x\left(S_n^xS_{n+1}^x+S_n^yS_{n+1}^y
\right)+J_zS_n^zS_{n+1}^z+\mu_BBS_n^z~.
\label{h_n}
\end{equation}

 $\mu_B$ is the Bohr magneton and the lattice spacing
will be denoted by $c$ ;
the $S_n$ are supposed to be  dimensionless.

Instead of the electrical current we consider the
magnetic current\cite{Shastry} 
\begin{eqnarray}
J_M& =& i\mu_B\frac{c}{\hbar}\frac{J_x}{2}
\sum_n\,\bigl[S_n^+S_{n+1}^--S_n^-S_{n+1}^+\bigr] 
\label{J_M}\\
& =& J_x\mu_B\frac{c}{\hbar}\sum_n\,\bigl[{\bf S}_n\times 
{\bf S}_{n+1}\bigr]_z
\ \equiv\ i\,\frac{c}{\hbar}[H,P_M]~,
\nonumber
\end{eqnarray}
where we have defined the magnetic polarization
$P_M =\mu_B \sum_n\,nS_n^z$, which is possible only
for chains with open boundary conditions (OBC).

For the energy current \cite{NieVia71,ZNP}
we have equivalently
\begin{eqnarray}
\frac{\hbar}{c}J_E& =& i\sum_n \left[h_n,h_{n+1}\right] \ = \
B\frac{\hbar}{c}J_M+\sum_n\tilde {\bf S}_n\cdot({\bf S}_{n+1}\times \tilde
{\bf S}_{n+2})
\nonumber\\
&=&B\frac{\hbar}{c}J_M+\sum_n\det\left[\tilde {\bf S}_n {\bf S}_{n+1}
 \tilde {\bf S}_{n+2}\right] \ \equiv\ i[H,P_E]~,
\label{J_E}
\end{eqnarray}
where we have used ${\bf S}_n=(S_n^x,S_n^y,S_n^z)$ and 
the definition
$\tilde {\bf S}_n=(J_xS_n^x,J_yS_n^y,J_zS_n^z)$.
In the following we will denote
with $j_i\equiv J_i/{\rm Vol},\;i\in\{M,E\}$
the respective current densities. Here
${\rm Vol}=cN$ is the one-dimensional volume,
where $c$ is the lattice constant and $N$ the
number of sites.

The energy polarization $P_E=\sum_n\,nh_n$ entering
Eq.\ (\ref{J_E}) for OBC represents a temperature
gradient, and, as is well known, \cite{Barnard} entails always
a magnetic polarization $P_M$,
which stems from the chemical- potential-term (magnetic- field-term).
To see this we insert a site-dependent (linear) 
$$\beta=\beta(n)=\bar \beta+nc\nabla\beta$$ 
into the
thermodynamical expectation value. \cite{ZMR} 
The Boltzmann-factor can be rewritten as follows:
\begin{eqnarray}
\exp\left(-\sum_n\beta(n)h_n\right)&=&\exp\left[-\bar\beta \left(
H+\frac{c\nabla\beta}{\bar\beta}\sum_nnh_n\right)\right]\nonumber\\
&=&\exp\left[-\bar\beta(H+FcP_E)\right]\label{def_P_E}
\end{eqnarray} 
where $F=\nabla\beta/\bar\beta=-(\nabla T)/\bar T$. 
This motivates the use of $P_E$ to model a temperature gradient.

\section{Correlation functions} \label{sec_corr_func}

Setting up the notation for the response theory
we define with\cite{SWZ,ZMR}
\begin{equation}
\Lambda(AB)(z) \ \equiv\
\frac{i}{\hbar}\int_0^\infty e^{izt}\bigl\langle [A(t),B]\bigr\rangle\,dt
\label{def_lambda}
\end{equation}
the retarded Green's function of two operators $A$
and $B$. In our case $A$ and $B$ will be mostly current operators;
we therefore introduce the notation 
$\Lambda(J_iJ_j)\equiv\Lambda_{ij}$. 
$\Lambda(AB)(z) $  vanishes when
one of the operators is a constant of motion and commutes
with the Boltzmann-factor:
by the cyclic property of the trace one has in this case
$\langle AB\rangle=\langle BA\rangle$.

The energy-current operator $J_E$ commutes with the 
Hamiltonian in the absence of an external magnetic field
$B$. Using Eq.\ (\ref{J_E}) we have
$\Lambda_{\mathit ME}=\Lambda_{\mathit EM}=B\Lambda_{\mathit MM}$
and $\Lambda_{\mathit EE}=B^2\Lambda_{\mathit MM}$.

The isothermal susceptibility\cite{ZMR} is given by
\begin{equation}
\chi^T(AB)\ \equiv\ \int_0^\beta \langle
\Delta A(\tau)\Delta B\rangle\,d\tau~,
\label{def_chi}
\end{equation}
with $\Delta A=A-\langle A\rangle$
and $\Delta B=B-\langle B\rangle$. It
takes the usual form
$\chi^T(AB)\ = \beta \langle
\Delta A\Delta B\rangle$
when one of the operators is a constant of motion.
The generalized Drude weight (which is used, e.g., in Ref.\
 \onlinecite{AlGr_kappa}) is defined by
\begin{equation}
\langle\langle AB\rangle\rangle \ \equiv\ 
\lim_{z\to 0}(-iz)\int_0^\infty
e^{itz}\langle \Delta A(t)\Delta B\rangle\, dt~.
\label{<<_>>}
\end{equation}
These three correlation functions are not independent,
comparing the respective
eigenstates-representation, e.g., 
$Z\Lambda(AB)(0)=\sum_{E_n\ne E_k} \Delta A_{nk}\Delta B_{kn}
\left(e^{-\beta E_k}-e^{-\beta E_n}\right)/\left(E_n-E_k\right)$,
 where $Z$ is the partition function, one finds
\begin{equation}
\beta\,\langle\langle A B \rangle\rangle\,+\,\Lambda(AB)(0)
\ =\ \chi^T(AB)~.
\label{relation}
\end{equation}
Under open boundary conditions, when the relations
$J_j=i\frac{c}{\hbar}[H,P_j]$ are valid, we can express the 
isothermal susceptibility,  Eq.\ (\ref{def_chi}),
\begin{equation}
\chi^{T,OBC}(J_iJ_j)\ =\ i\frac{c}{\hbar}\bigl\langle[J_i,P_j]\bigr\rangle
\label{rel_obc}
\end{equation}
as a static expectation value. Note, however,
that the value of the susceptibility is
independent of the boundary conditions in the
thermodynamic limit: $\chi^{T,OBC}(J_iJ_j) =\chi^{T,PBC}(J_iJ_j)$.

\section{Magnetothermal coefficients} \label{sec_coeff}

We now discuss the general recipe for
 the computation of the entries of the
$L$-Tensor, appearing in Eq.\ (\ref{L_tau}).

 
We assume a perturbation in the form of a
polarization, (cf., e.g., Ref. \onlinecite{ZMR}) i.e., we add to the Hamiltonian a term
 $c\nabla B\cdot P_M$ 
with $P_M:=\mu_B\sum_n nS_n^z$ and 
compute the response of the current
density operator to obtain the MM-component.
(A different but equivalent approach is given in Ref.\ \onlinecite{SWZ}.)
To compute the 
remaining entries of the $L$-Tensor, we replace the magnetic by the
energy current density (in the second row) and/or substitute $P_E:=\sum_n
nh_n$ for $P_M$ -- as well as $c\nabla T$ for $c\nabla B$   --
 and add a factor $k_B\beta$
 (in the second column). [One should compare this procedure with Eq.\ (\ref{def_P_E}).]

The linear response theory gives 
the following contribution to $\hat L_{ij}$:\cite{ZMR}
$$
(k_B\beta)^{j-1}\frac{c}{\hbar}
\int_0^\infty e^{izt}\bigl\langle i[j_i,P_j(-t)]\bigr\rangle\, dt
$$
where $\beta ^M:=\beta$, $\beta^E:=\beta^2$, etc.,
and $j_i=J_i/{\rm Vol}$ (see Sec.\ \ref{sec_operators}).
Integration by parts yields for the above expression 
\begin{displaymath}
\lim_{z\to 0} {(k_B\beta)^{j-1}\over -i z}
\left[i\frac{c}{\hbar}
\bigl\langle[j_i,P_j]\bigr\rangle-
\underbrace{\frac{i}{\hbar}\int_0^\infty e^{izt}
\bigl\langle [j_i(t),J_j]\bigr\rangle\, dt}_
{\Lambda_{ij}(z) }\right]~.
\end{displaymath}
One should note that $\Lambda_{ij}(0)=0$ if one
of the currents commutes with the Hamiltonian, and consequently there
is no regular part.
The factor $(-iz)^{-1}$ accounts for the delta-function
in the zero-frequency limit and we have
\begin{equation}
L_{ij}\ = \ (k_B\beta)^{j-1}\left\{
i \frac{c}{\hbar}
\bigl\langle[j_i,P_j]\bigr\rangle- \Lambda_{ij}(0)\right\}~.
\label{L_ij}
\end{equation}
It is also possible to express this result
by the correlation function defined in Eq.\ (\ref{<<_>>})
using Eqs.\ (\ref{relation}) and (\ref{rel_obc}):
\begin{equation}
L_{ij}\ =\ (k_B\beta)^{j-1}\left[
\chi^T(j_iJ_j) -\Lambda_{ij}(0) 
                      \right] \ =\ k_B^{j-1}
\beta^j\,\langle\langle j_iJ_j\rangle\rangle~.
\label{L_<<>>}
\end{equation}
From Eq.\ (\ref{L_ij}) follows the well-known result for
the magnetic Drude weight
$L_{\mathit MM}\ =\ \langle -T_{\mathit MM}\rangle  \frac{c}{\hbar}-
\Lambda_{\mathit MM}$,
where $T_{\mathit MM}=-i[j_M,P_M]$
is the kinetic energy per site -- apart from a prefactor.
In view of Eq.\ (\ref{L_<<>>}) we define the {\em generalized kinetic energy}
 \begin{equation}\label{genkin}T_{\mathit ij}=-i[j_i(B=0),P_j(B=0)].\end{equation}
Here, one should note a particularity  produced by the
twofold r\^ole played by the operator $P_E$ in the xxz-chain:
it not only describes thermal response but also acts as a `boost'-operator for
the constants of motion.\cite{GRMA} Hence, $J_E$ as well as $T_{EE}$ are constants
of motion.
 
Taking into account that $J_E$ commutes with the
Hamiltonian for $B=0$ and using Eqs.\ (\ref{J_E})
and (\ref{L_ij}) we have
\begin{eqnarray}
L_{\mathit EM}& =& i\, \frac{c}{\hbar}
\langle [j_E,P_M]\rangle - B\Lambda_{\mathit MM} 
\nonumber \\
&=& i\, \frac{c}{\hbar}\bigl\langle [j_E(B=0),P_M]\bigr\rangle
+B\left\{ i \frac{c}{\hbar}\bigl\langle [j_M,P_M]\bigr\rangle - \Lambda_{\mathit MM}
  \right\}
\nonumber\\
&\equiv&\langle -T_{\mathit EM}\rangle\frac{c}{\hbar}+BL_{\mathit MM}
\label{L_MMEM} 
\end{eqnarray}
and at finite $B$: 
$\hbar L_{\mathit EE}/(ck_B\beta)\ =\  
\{i\langle[j_E(B=0)+Bj_M,P_E(B=0)+BP_M]\rangle-B^2\Lambda_{\mathit MM}\}=
B\langle -T_{\mathit ME}\rangle
+ B\langle -T_{\mathit EM}\rangle
- \langle T_{\mathit EE}\rangle - B^2\{
\langle T_{\mathit MM}\rangle+
\hbar\Lambda_{\mathit MM}/c\}$.

The generalized kinetic energy $T_{\mathit EM}=-i[j_E,P_M]$ 
appearing in Eq.\ (\ref{L_MMEM}) is given by 
(in units $\mu_Bc/\hbar$)
\begin{eqnarray} T_{\mathit EM}&=&
\frac{-i}{\rm Vol}\sum_n\Bigl(n\epsilon_{\alpha\beta\gamma}
\bigl[\tilde { S}_n^\alpha, S_n^z\bigr] 
S_{n+1}^\beta\tilde S_{n+2}^\gamma+{\rm cycl.}\Bigr) \nonumber\\
&=&\sum_n\left\{n\tilde {\bf S}_{n}\times( {\bf S}_{n+1}\times
 \tilde {\bf S}_{n+2})+{\rm cycl.}\right\}_z/{\rm Vol}\nonumber\\
&=&\frac{1}{\rm Vol}\sum_n\Bigl\{\tilde {\bf S}_{n+2}\times(\tilde {\bf S}_n\!\times
 {\bf S}_{n+1})\nonumber\\
&&-\tilde {\bf S}_n\times({\bf S}_{n+1}\times
\tilde {\bf S}_{n+2})\Bigr\}_z\nonumber\\ 
&=& {J_x^2\over {\rm Vol}}
\sum_n\ \Bigl\{-S_n^+S_{n+1}^zS_{n+2}^--S_n^-S_{n+1}^zS_{n+2}^+\nonumber\\
&&+\frac{J_z}{2J_x}\bigl(S_n^+S_{n+1}^-S_{n+2}^z+S_n^-S_{n+1}^+S_{n+2}^z\nonumber\\
&&+S_n^zS_{n+1}^-S_{n+2}^++S_n^zS_{n+1}^+S_{n+2}^-\bigr)\Bigr\} ~,
\label{temform}
\end{eqnarray}
where `cycl.' means cyclic permutation of $\{n,n+1,n+2\}$ --
note that the $\tilde{\phantom a}$-accent is permuted with the indices.

$\langle -T_{\mathit EM}\rangle$ changes sign under
inversion of the $S^z$-magnetization $M$ and vanishes consequently
for $B=0$, as does the magnetothermal response in this case,
compare (\ref{L_MMEM}).
The case of a finite $B$ is therefore of interest. 
We will consider only the effect linear in $B$, since
magnetic fields amount normally only to small energies: 
\begin{equation}\label{lem}
L_{EM}\approx
B\frac{\partial}{\partial B}L_{\mathit EM}\Big|_{B=0}
=B\Bigl[\bigl\langle -T_{\mathit EM}(-\beta M)\bigr\rangle\frac{c}{\hbar}
+L_{\mathit MM}\Bigr]~.
\end{equation}
The  EM-coefficient, therefore, consists of two parts. One is
the $B$-derivative of the zero-field $L_{\mathit EM}$;
 the other comes from the change of $J_E$ in magnetic field
(namely, it acquires a term $BJ$) and  is already 
linear in $B$ -- for small $B$.

The quotient of $L_{ME}$ and $L_{MM}$
\begin{equation}
Q_M\ =\ \hat L_{\mathit ME}/\hat L_{\mathit MM}
\ =\ L_{\mathit ME}/ L_{\mathit MM}~. 
\label{Q}
\end{equation}
is called the {\em (magnetic) thermopower}. It has the advantage
that geometric properties of samples cancel in
experimental studies and that the relaxation times cancel.

To discuss $L_{\mathit ME}$ is superfluous, if one keeps
the Onsager relation in mind. Anyway, if we follow the same steps as above,
we arrive -- with explicitly using OBC -- at (in units $\mu_Bc/\hbar$)
\begin{eqnarray*}T_{\mathit ME}&=&-i[j_M,P_E]\\
&=&\frac{-i}{\rm Vol}\sum_n\Bigl( n\left[\tilde{\bf S}_n\times{\bf
 S}_{n+1},\tilde{\bf S}_n\cdot {\bf S}_{n+1}\right]
+\\
&&(n+1)\tilde {\bf S}_n\times\left[{\bf S}_{n+1},{\bf S}_{n+1}\cdot
\tilde {\bf S}_{n+2}\right]+\\
&&n\left[{\bf S}_{n+1},\tilde{\bf S}_{n}\cdot
 {\bf S}_{n+1}\right]\times \tilde {\bf S}_{n+2}\Bigr)_z\\ 
&=&\sum \bigl(n/2J_x^2({\bf S}_n-{\bf S}_{n+1})-
(n+1)\tilde {\bf S}_n\times({\bf S}_{n+1}\times \tilde {\bf S}_{n+2})\\
&&-n\tilde {\bf S}_{n+2}\times(\tilde {\bf S}_n\times
 {\bf S}_{n+1})\bigr)_z/{\rm Vol}\\
&=&\sum_nJ_x^2/2\Bigl\{
-S_n^+S_{n+1}^zS_{n+2}^-
-S_n^-S_{n+1}^zS_{n+2}^+\\
&&+(n+1)J_z/J_x\bigl(S_n^+S_{n+1}^-S_{n+2}^z
+S_n^-S_{n+1}^+S_{n+2}^z\bigr)\\
&&-nJ_z/J_x\bigl(S_n^zS_{n+1}^-S_{n+2}^+
+S_n^zS_{n+1}^+S_{n+2}^-\bigr)\\
&&+M/\mu_B- S_0^z\cdot {\rm Vol}/c\Bigr\}/{\rm Vol},\end{eqnarray*}
which differs apparently from $ T_{\mathit EM}$. This does not contradict
the Onsager relation because we find that
  $\langle T_{ME}
\rangle = \langle T_{\mathit EM}\rangle$ still holds.
This may be seen by first using Eqs. (\ref{J_M}) and (\ref{J_E}):
$$\langle T_{ME}\rangle-\langle T_{EM}\rangle
=\frac{c}{\hbar}\Bigl\langle\bigl[[H,P_M],P_E\bigr]-
\bigl[[H,P_E],P_M\bigr]\Bigr\rangle$$
then adding a term which is zero by the cyclic property of the trace:
$$=\frac{c}{\hbar}\Bigl\langle\bigl[[H,P_M],P_E\bigr]+\bigl[[P_M,P_E],H\bigr]
+\bigl[[P_E,H],P_M\bigr]\Bigr\rangle.$$
That this final expression is zero is just Jacobi's identity.

Here it should be emphasized that $\langle T_{ME}\rangle$ has  non-negligible
contributions from the boundary, and is
-- unlike $\langle T_{\mathit EM}\rangle$ --  sensitive to a change from PBC to OBC.

\section{Jordan-Wigner transform and free fermion model}
     \label{sec_jordan}

For our spin Hamiltonian all 
quantities may easily be  calculated when $J_z=0$
via the Jordan-Wigner mapping to a free fermion system.
Under this condition the Hamiltonian is straightforwardly diagonalized
by a Fourier transform with eigenvalues $\cos k$
(setting $J_x\equiv1$)
following the Fermi-Dirac distribution
$\langle n_k\rangle=\langle c^\dag_kc_k^{\phantom\dag}\rangle=
\bigl[1+\exp\bigl(\beta \cos k \bigr)\bigr]^{-1}$. 

The Drude weight entering (\ref{lem}) is given simply by
$-L_{MM}=\langle T_{MM}\rangle c/\hbar
=\int\cos k\langle n_k\rangle dk/(2\pi c)(\mu_B^2c/\hbar ^2)$.
At the same time $T_{\mathit EM}$ simplifies to 
(again in units $\mu_Bc/\hbar$): 
\begin{eqnarray*}T_{\mathit EM}&=&{1\over {\rm Vol}} \sum_n\left\{
-S_n^+S_{n+1}^zS_{n+2}^-
-S_n^-S_{n+1}^zS_{n+2}^+\right\}\\
&=& -\int {dk\over 2\pi c} \cos(2k) n_k ~.
\end{eqnarray*}
Following (\ref{lem}) we need to compute --right-hand side in units
$\mu_B^2c/\hbar$ --
$$\bigl\langle (-\beta M)(-T_{\mathit EM})
\bigr\rangle\ =\
{-\beta\over {\rm Vol}}\, \sum_{k,q}\,\cos(2k)
\left\langle n_k \left(n_q-\frac{1}{2}\right)\right\rangle~.
$$
$\langle n_kn_q\rangle =\langle n_k\rangle\langle n_q\rangle$
holds for $k\neq q$, $n_k^2=n_k$ and 
$\sum_{k,q}\cos(2k)\ \langle n_k\rangle\langle n_q\rangle=0$.
We find:
\begin{eqnarray} \nonumber
\bigl\langle (-\beta M)(-T_{\mathit EM})\bigr\rangle&=&
{\beta\over{\rm Vol} }
\sum_k\bigl[-\cos(2k)\bigr]\bigl[\langle n_k \rangle-
\langle n_k\rangle^2\bigr]\\
&=&-\beta\int \frac{dk}{2\pi c}\frac{\cos(2k)}
{4\cosh^2\bigl[\beta \cos(k)/2\bigr]}~.
\label{T_EM_xy}
\end{eqnarray}
This result can be related with the temperature-derivative
of the kinetic energy, 
\begin{equation}\label{freefermion}
\bigl\langle T_{\mathit EM}(-\beta M)\bigr\rangle
\ = \ \frac{d}{dT}\bigl[T\langle
-T_{\mathit MM}\rangle\bigr]~,
\end{equation}
where 
$T_{\mathit MM}$ for the xy-Model is defined above.
The relation (\ref{freefermion}) is easily verified 
using a partial integration.

\section{Einstein relations} \label{sec_einstein}

Let us recapitulate Einstein's relation for
diffusive transport in a metal:
In a closed system the {\em diffusive} current 
$j_D=-\hat D\nabla n$, driven by a gradient in the 
particle density $n$, and the electrical current
driven by an {\it external potential},
$j=\hat\sigma E$, add to zero: $j_D=-j$. 
This condition yields Einstein's famous relation:
\begin{equation}
\frac{\hat\sigma}{\hat D}\ =\ \frac{\nabla n}{E}~.
\label{el_einstein}
\end{equation}

Here, in the case of  a magnetic system,
we are not interested in a variation of the electron
particle density, but in a change in internal energy or magnetization.
Hence, we replace $n(x)\to A_n$ where $A_n$ is either $h_n$ or $\mu_BS_n^z$.
Furthermore, the currents are not driven by a field $E$, but
by a polarization $P_F=\sum_n nF_n$ (times the lattice constant $c$), for a
conserved $F$: $[F,H]=0$. In our case the operator $F$ is
either the magnetization or the internal energy.

We will also assume that there is diffusive transport for both the
magnetization and the energy current. Thereby, we are led to the
following definitions of the corresponding diffusion coefficients
$$j_M\equiv\hat D_{MM}\nabla M\qquad j_E\equiv \hat D_{EE}\nabla E $$
where $M$ and $E$ are the magnetization and internal energy.

Generalized Einstein relations can be derived using
{\em static response} of $\nabla A|_n:=(A_{n+1}-A_n)/c$ 
to a  perturbation $cP_F$, where $c$ is
the lattice constant along the chain:
$$\nabla n/E\ \rightarrow\ \sum_n\chi^T(\nabla A|_ncP_F)/{\rm Vol}~,
$$
where we perform a volume average. (This becomes mandatory
 because we consider the response to the current density rather than
to the current at a fixed site.)
 Using the linearity of the isothermal
susceptibility we find
(by a discrete version of an  integration by parts):
\begin{eqnarray*}
c\,\chi^T(\nabla A|_nP_F)&=& \sum_m\Bigl\{
\chi^T\bigl[A_{n+1}(mF_m)\bigr]- \chi^T\bigl[A_n(mF_m)\bigr]
                               \Bigr\}
\\
&=&\sum_m\Bigl[
(m+1)\chi^T\bigl(A_{n}F_{m}\bigr)-m\chi^T\bigl(A_nF_m\bigr)
                               \Bigr] \\
&=&\sum_m\chi^T(A_nF_m)\end{eqnarray*}
and with the volume average on both sides 
\begin{equation}\sum_n c\,\chi^T(\nabla A|_nP_F)/{\rm Vol}=\beta
 \langle \Delta A\Delta F\rangle/{\rm Vol}~.
\label{intbypartII}\end{equation}

We rewrite (\ref{el_einstein}) for the case of 
magnetothermal response, 
$j_M\equiv \hat L_{\mathit ME}\nabla T=
 \hat D_{\mathit MM}\nabla M\equiv -j_D$,
and find
\begin{eqnarray}
\label{derivation_1}
{\hat L_{ME}\over \hat D_{MM}} & =&
{L_{ME}\over D_{MM}} \ =\
{\nabla M\over \nabla T} \ =\ 
{k_B\over k_BT}{\nabla M\over (\nabla T/T)} \\
&=& \nonumber
k_B\beta\,\chi^T\left(\nabla M|_n P_E\right) \ =\ 
{k_B\beta^2\over {\rm Vol}}\langle \Delta M \Delta H\rangle~,
\end{eqnarray}
where we have assumed equal relaxation times $\tau$
for $\hat L_{ME}=\tau L_{ME}$ and the magnetization
diffusion: $\hat D_{MM}=\tau D_{MM}$.
(This assumption was only made for simplicity.
 At least, to our knowledge
there is no reason why the relaxation times should equal each other.
However, the assumption is natural as in both cases the finite
relaxation time comes from the fact that magnons lose momentum, and the
physical processes responsible for that should be in both cases the same.)
Eq.\ (\ref{derivation_1}) leads to 
\begin{eqnarray}\nonumber
{L_{\mathit ME} \over D_{\mathit MM}}& =& 
\frac{k_B\beta^2}{{\rm Vol}}\langle\Delta M \Delta H\rangle
\ \approx\ B{d\over dB}
\frac{k_B\beta^2}{{\rm Vol}}\langle\Delta M \Delta H\rangle\\
&=&
{Bk_B\beta^2\over {\rm Vol}}\langle\Delta M \Delta M\rangle
-{Bk_B\beta^3\over {\rm Vol}}\langle\Delta M^2 \Delta H\rangle
\label{deriv_line2}
\end{eqnarray}
for small magnetic fields $B$, the structure of this equation is similar to Eq.\ (\ref{lem}).
The first term of the right-hand-side of Eq.\ (\ref{deriv_line2})
is just $B\,\chi/T$, where 
$\chi =\beta\langle \Delta M^2\rangle/{\rm Vol}$ is the
magnetic susceptibility. The second 
term of the right-hand-side of Eq.\ (\ref{deriv_line2})
simplifies to
\[
\frac{Bk_B\beta^3}{{\rm Vol}}\frac{d}{d\beta}\langle \Delta M\Delta M
\rangle
= {k_B B\beta ^3}\frac{d}{d\beta}{\chi\over \beta}
=-{B\over T }\frac{d}{d T}[T\chi] 
\]
We therefore find the new relation:
\begin{equation}
{L_{\mathit ME} \over D_{\mathit MM}}\ =\  {B\over T}\chi
-{B\over T }\frac{d}{d T}[T\chi]
\ =\ -B\frac{d\, \chi}{d T}~.
\label{ME_chi}
\end{equation}
Classically, the diffusion constant is $\hat D=v^2\tau$,
where $v$ is the velocity of the elementary excitations,
here the magnon velocity. This leads at low-temperatures
to a temperature-independent diffusion coefficient
$D=\hat D/\tau = v^2$ and via (\ref{ME_chi}) to 
vanishing magnetothermal response for $T\to0$, whenever
$\chi(T)$ becomes constant for $T\to0$.
Setting $A=M$ and $F=M$
in the general Einstein relation
yields \cite{ChLu,AlGr_MC}
\begin{equation}
L_{MM}/D_{MM}\  =\  \chi~,
\label{MM_chi}
\end{equation}
and with the choice $A=H=F$ one may obtain \cite{ChLu,KlSa}
an analogous relation:
$L_{\mathit EE}/D_{\mathit EE}= c_V$.

We will now introduce a formula which 
we believe to be approximately valid at small $T$.
As stated above, the diffusion constant is known not to vary much 
near $T=0$. Because of our restriction to low $T$ we
assume a temperature-independent diffusion coefficient
$D_{MM}$ (see discussion above).
Then we can  rewrite (\ref{MM_chi}) as
$\frac{d}{dT}L_{\mathit MM}=\frac{d}{dT}[\chi D_{MM}]
\approx D_{MM}\frac{d}{dT}\,\chi $.
Inserting this expression into (\ref{ME_chi}) we obtain
\begin{equation}\label{newform}
(-L_{\mathit ME})\ =\
B\frac{d}{dT}L_{\mathit MM}.
\end{equation}
In the case of free fermions this is just the result
of Eq.\ (\ref{freefermion}).
In the case of a finite interaction it seems to be correct in the
limit $T\to 0$ if we use data from Ref.\ \onlinecite{Kl_Drude}.
We provide tests of this relation in section \ref{sec_num_res}.
\section{An exact Einstein equation}
\label{sec_exact_Einstein}


In Sec.\ref{sec_einstein} we did set-up several 
versions of generalized Einstein relations
appropriate for magnetothermal response.
Those relations may be viewed as a link 
between the dynamical and static response theory,
as they connect corresponding correlation functions by
introducing diffusion constants.

Upon closer examination of the right-hand sides of the Einstein
relations like (\ref{MM_chi})-- namely,
$\chi$, $c_V$ and $d\chi/dT$ -- we find that
that these are static expectation values 
of products of $\Delta M$- and $\Delta H$-operators 
which could be generated by derivatives.
Hence, it is an easy task to establish functional 
relations between the static correlations,  e.g.,
$$
T^2\frac{d^2}{dB^2}c_v\ =\ \frac{d}{dT}\left[
T^2{d\over dT}\big(T\chi\big) \right].
$$ 
An intriguing question is,
whether the analogous equations obtained by switching between static and
dynamical responses --  $\chi\leftrightarrow L_{MM}$, etc.  -- could also be valid. For one of these relations,
Eq.\ (\ref{newform}), 
the range of validity is confined to the low
temperature regime unless $J_z=0$. (A detailed discussion follows in
Sec.\ \ref{sec_num_res}.)
One reason for the failure could be the fact that in this particular
case one of the correlation functions does not reduce to a thermodynamical 
expectation value.
We therefore focus on a relation between the thermal and magnetothermal
response coefficients -- both mere thermal expectation values.
We claim that the relation
\begin{equation}
T^2\frac{d^2}{dB^2}\langle -T_{EE}\rangle
\ =\ \frac{d}{dT}\left[T^2\langle T_{EM}
(-\beta M)\rangle \right]
\label{exact_ein}
\end{equation}
is exact.
The proof is provided in the appendix.
This equation could be useful to derive
an analytical solution for the magnetothermal response.

  The arguments of the proof
 rely mainly on the fact that one of the currents is
a constant of motion. It is by the same line of arguments possible
 to show
that Eq. (\ref{newform}) is exact in the case
where $[H,J_M]=0$. Unfortunately, this is a very restricting
constraint; only the  xy-model and the Haldane-Shastry
Hamiltonian\cite {HalShas} meet
the requirement. 

\section{Numerical results} \label{sec_num_res}

In the free fermion case $L_{EM}$ is easily accessible by a
simple  evaluation of the analytic expression
(\ref{T_EM_xy}) and a corresponding
one for $L_{MM}=-\langle T_{MM}\rangle$,
using (\ref{lem}).
If $J_z\neq 0$ we have to compute a mere thermal 
expectation value in order to obtain 
$T_{EM}$, a task which
is tractable to Monte-Carlo-simulations.
Here, we use the stochastic series expansion 
(SSE), which is a Taylor expansion 
based Monte-Carlo-method. \cite{San_loop}

We assume a Hamiltonian of the form $H=\sum_n J_nh_n$,
we may write the partition function 
$Z=\sum\frac{(-\beta)^m}{m!}
\langle\alpha|\prod_{i=1}^mJ_{\phi_m(i)}h_{\phi_m(i)}
|\alpha\rangle$
where  the sum runs over all numbers $m$, all functions 
$\phi_m:\{1,\dots,m\}\to {\mathbb N}$, and $S^z$-eigenbasis-states $\alpha$.
The SSE-program now samples the products of operators appearing 
in this expression of $Z$ with their relative weight factors.

Normally, with MC-methods it is problematic to measure operators
which are not diagonal in the $S^z$-eigenbasis. But with SSE
another class of operators is easily accessible:
 Following a standard procedure in statistical mechanics,
we have $\langle h_n\rangle=(\partial_{J_n}Z)/Z$.
For the SSE  this means that we can measure 
any operator $h_n$ appearing in the Hamiltonian simply by counting how often it
appears in the products sampled by our SSE-program (and dividing by $J_n$).
So it is fairly easy to measure sums of products of parts of the Hamiltonian as
$S_n^+S_{n+1}^-$, $S_n^-S_{n+1}^+$ or $S_n^zS_{n+1}^z$.
To make use of this fact in the given context, it is  expedient to
find an expression for $T_{EM}$ which consists only of terms appearing in the Hamiltonian.
Here we present one ($T_{EM}$ in units $\mu_B c/\hbar$):
\begin{eqnarray*}
T_{EM}&=&\frac{-i\hbar}{\mu_B c}\sum_n\{[j_n,h_{n+1}]+[h_n,j_{n+1}]\}/{\rm Vol}\\
&=&\sum_n\frac{1}{2}\Bigl\{J_x^2
\left[S_n^+S_{n+1}^--S_n^-S_{n+1}^+,S_{n+1}^+S_{n+2}^-\right]\\
&&+J_x^2
\left[S_n^+S_{n+1}^--S_n^-S_{n+1}^+,S_{n+1}^-
S_{n+2}^+\right]\\
&&+J_xJ_z\left[S_n^+S_{n+1}^--S_n^-S_{n+1}^+,S_{n+1}^zS_{n+2}^z\right]\\
&&+J_xJ_z\left[S_n^zS_{n+1}^z,S_{n+1}^+S_{n+2}^--S_{n+1}^-
S_{n+2}^+\right]\Bigr\}/{\rm Vol}~,
\end{eqnarray*} 
which then allows to assess the magnetothermal coefficient by the SSE.

The Drude weight -- which is the other
input in Eq.\ (\ref{lem}) -- may in principle be calculated by the
Bethe-Ansatz method. This was attempted by 
several authors, \cite{zot99,Kl_Drude} but
their calculations are still under discussion, \cite{AlGr_Drude} 
and so far there are no reliable results.
We therefore use exact-diagonalization to obtain data
for the Drude-peak. This has clearly the disadvantage that
we cannot make any statements about the low-$T$
and high-$J_z$ regimes where the convergence 
with ${\rm Vol}\to \infty$ is slow (cf. Ref. \onlinecite{NMA}).

\begin{figure}
\epsfig{file=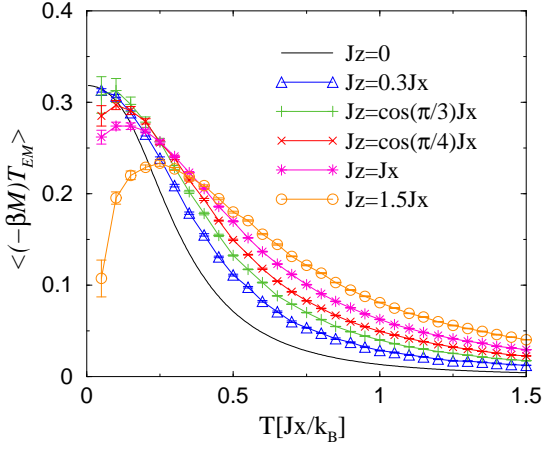,width=0.4\textwidth,angle=0}
\caption{Monte-Carlo-data (48 \& 96 sites, PBC) for the
         dimensionless $\langle (-\beta M)T_{\mathit EM}\rangle$ 
	 and various interaction strengths $J_z/J_{xx}$.
	 The statistical MC-error-bars are given.}
\label{figTem}
\end{figure}

\begin{figure}
\epsfig{file=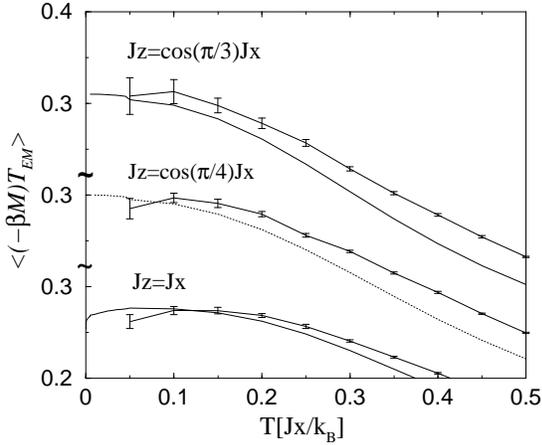,width=0.4\textwidth,angle=0}
\caption{Monte-Carlo-data (48 \& 96 sites) at small $T$ for PBC
         in comparison (lines) with the  
	 estimates for
	 $\langle (-\beta M)T_{\mathit EM}\rangle$ 
	 by Eq.\ (\ref{newform}) and data from Ref. \onlinecite{Kl_Drude}
         for the Drude weight. 
         The cases $J_z=\cos(\pi/4)$ and $J_z=1$ are offset 
	 for clarity.
	 }
\label{figcom}
\end{figure}

In Figs.\ \ref{figTem} and \ref{figcom}
we plot $\bigl\langle (-\beta M)T_{\mathit EM}\bigr\rangle$
in units of $J_x\mu_B^2/\hbar$ -- a detailed discussion of units follows in the
next section, compare Eqs.\ (\ref{dimen1}) and (\ref{L_MMEM}) --
as a function of temperature for various
interaction strengths. 
The data for $J_z\neq 0$ are obtained by MC-simulations
and -- in Fig.\ \ref{figcom} -- by formula
Eq.\ (\ref{newform}) and Eq.\ (\ref{lem}).

The MC-results in Fig.\ \ref{figTem} 
 are clearly not good enough to determine exactly
the $T=0 $- value.
However, for $J_z=J_x$ the data  seems to extrapolate to
$0.25$, which is the Bethe-Ansatz result for the dimensionless
$T=0$ Drude weight for the isotropic Heisenberg chain. \cite{Shastry}
This result would imply via (\ref{lem}) a vanishing
magnetothermal response for $T\to0$ and via (\ref{newform})
a vanishing $T$-derivative of the Drude weight, in contrast
to previous results. \cite{zot99}

In Fig.\ \ref{figcom} we present a comparison
(at low $T$) between
the MC-results for $\bigl\langle (-\beta M)T_{\mathit EM}\bigr\rangle$
and the results from equation
(\ref{newform}), where we have used
Bethe-Ansatz results (because of the finite size gap
exact diagonalization provides here no alternative) from
Ref.\ \onlinecite{Kl_Drude} -- 
which we prefer
to Ref.\ \onlinecite{zot99} because it is in agreement with  Ref.\ \onlinecite{AlGr_Drude} --
 for the temperature-dependent
Drude weight $L_{MM}$. The agreement is very good for $T<0.2$.
As a consequence we may deduce from
Eq.\ (\ref{newform}) that 
$\bigl\langle (-\beta M)T_{\mathit EM}\bigr\rangle=
L_{\mathit MM}\hbar/c$ at $T=0$, 
the linear magnetothermal effect as
given by Eq.\ (\ref{lem}) vanishes consequently
for $T\to0$.

\begin{figure}
\epsfig{file=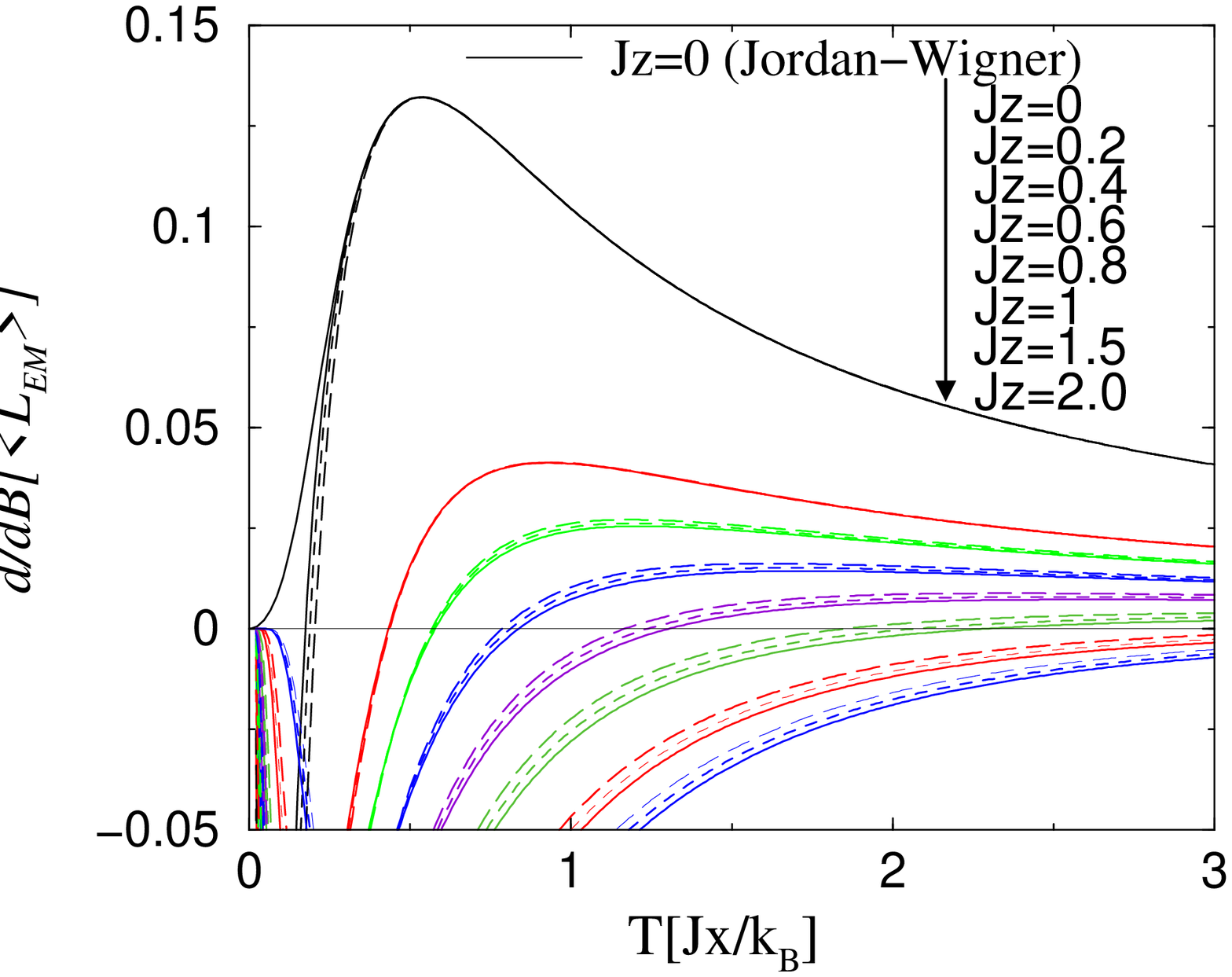,width=0.4\textwidth,angle=0}
\caption{The linear part (in $B$) of  
         $L_{\mathit EM}$ as a function of temperature,
         see Eq.\ (\ref{lem}), obtained by exact diagonalization
         for both
         $\langle (-\beta M)T_{\mathit EM}\rangle$ 
          and the Drude weight. We plotted data for three
          system sizes (14, 16, 18 sites) with different line styles
          (long-dashed, dashed, solid, respectively).
	 }
\label{figlem}
\end{figure}

\begin{figure}
\epsfig{file=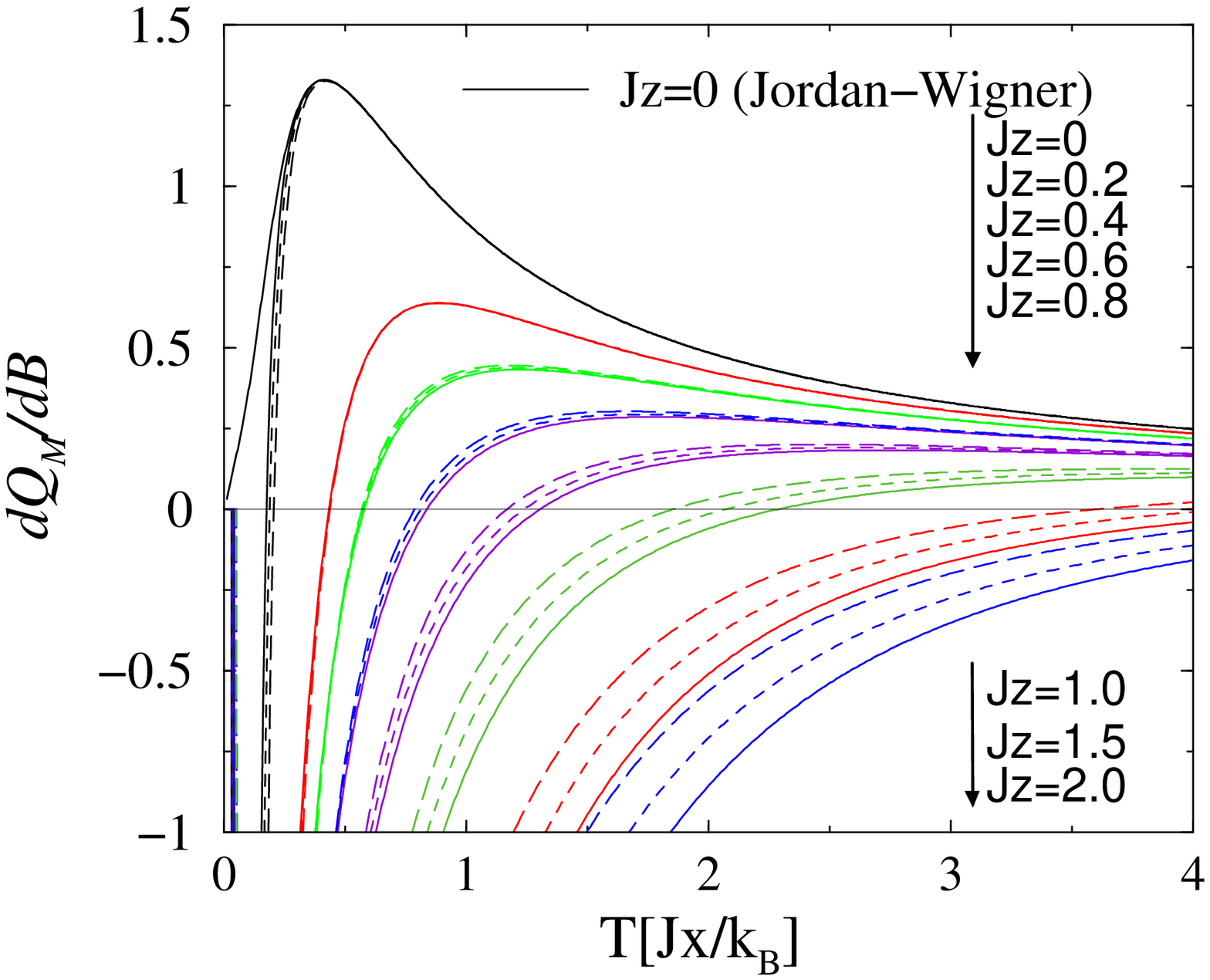,width=0.4\textwidth,angle=0}
\caption{The part of the thermopower (\ref{Q}) linear in $B$
         as a function of temperature,
         see Eq.\ (\ref{lem}) with exact-diagonalization-data
         for $\langle (-\beta M)T_{\mathit EM}\rangle$ and 
         the Drude weight. We plotted data for three
          system sizes (14, 16, 18 sites) with different line styles
          (long-dashed, dashed,  solid, respectively).}
\label{figQM}
\end{figure}

The results for $L_{\mathit EM}$ and  the
thermopower (the prime denotes a derivative with respect to the magnetic
field $B$)
$Q^\prime_M=L^\prime_{\mathit ME}/L_{\mathit MM}$ 
are 
presented in Fig.\ \ref{figlem} and Fig.\ \ref{figQM}, 
respectively.
Here we use exact-diagonalization, because for the computation of
the Drude weight QMC-methods
normally fail at higher $T$. 
(The
standard proceeding is to extrapolate
 from the Matsubara
frequencies to $\omega=0$ as attempted in Ref. \onlinecite{AlGr_MC}.
 This procedure becomes soon unstable if $T$ -- and
hence the spacing of the Matsubara frequencies -- grows.)
 We use exact-diagonalization for the computation of $\langle
 T_{EM}\rangle$
as well. 
Since $\langle
 T_{EM}\rangle$ converges much faster
with system size than the Drude weight, the finite size error
is determined by the Drude weight, so 
using QMC for $\langle
 T_{EM}\rangle$ alone  would not give better results.
On the contrary, if we used MC-data we would introduce the
statistical  error.
For the Drude peak we use Eq.\ (\ref{L_<<>>})
with $\langle\langle j_iJ_j\rangle\rangle=
\sum_{E_n= E_k} \langle n|j_i|k\rangle\langle k|J_j|n\rangle
e^{-\beta E_n}/Z$ ($Z$ is again the partition function)
for $\langle T_{EM}\rangle$ we simply use the expression
in Eq. (\ref{temform}).
The use of these expressions makes it possible for us to exploit
translational symmetry which allows us
to consider slightly larger systems than in Ref. \onlinecite{NMA} 
(namely, up to 18 sites).
 
Because of the resulting finite size gap we can only discuss
the high-$T$ regime. (In the xy-model we find a
vanishing thermopower at $T=0$ with a linear power-law.)

For small $J_z$
we find a maximum
at intermediate temperatures which  decreases
 and is shifted to higher $T$ as we increase $J_z$.

At higher $J_z$ the convergence becomes slower, so we cannot
make any precise statements about the transport coefficients. However,
 in Fig.\ \ref{figlem} and Fig.\ \ref{figQM}
the curves decrease with system size (for $T$ sufficiently large).
 This behavior can
be confirmed for all systems sizes (smaller than 18 sites).
If we assume that the direction of convergence does not change for larger
system sizes, our curves provide an upper bound
 to the exact results .
But this would imply a reversion
of the effect (negative thermopower).
While at $J_z=0$ the effect is always positive
-- apart from a reversion due to the finite size gap, one finds
$Q^\prime_M\leq 0$
 for a certain $T$-interval if $J_z$
is large enough. 
Because of the slow
convergence it is impossible for us to predict the precise location of
this interval. Here one should emphasize that our conclusions were drawn
from the inspection of relatively small systems, so we cannot really
exclude that the observed reversion is only due to finite size effects.

However, we see the reason for this reversion in the fact that the
 magnetothermal
coefficient consists of two summands [see Eq. (\ref{lem})] which turn out
to have opposite signs.
These behave very differently under a change of $J_z$.
While the Drude weight dominates at $J_z=0$, it drops dramatically as
$J_z$ is increased towards 1 (this was also reported in
Ref. \onlinecite{zot99}).
 The generalized kinetic
energy does not decay so much, such that it becomes the dominating part.

\section{Size of the magnetothermal effect} \label{sec_size}

Based on Eqs.\ (\ref{Ltensor}) and (\ref{dimen}) and
using the low $B$ approximation in Eq.\ (\ref{lem})
we present in this section estimates of the size of
the magnetic Seebeck effect.
\subsection{Dimensional analysis} \label{sec_dim_analysis}
As we are considering effects linear in
the magnetic field $B$ we use
$\prime$ as a shorthand for $\frac{\partial}{\partial B}$. 
Using $[J_M]=[\mu_B]{\rm m/s}$ and $[J_E]={\rm Wm}$ 
(compare Eq.\ (\ref{J_M}) and (\ref{J_E}) and
[$j_{M/E}]=[J_{M/E}]/{\rm m}$) we may decompose
the response coefficients into dimension-full and
dimension-less quantities \cite{AlGr_kappa}:
\begin{eqnarray}\label{dimen}
L^\prime_{\mathit ME}\ &=&\
\frac{k_B\mu_B}{\hbar}\frac{\mu_B c}{\hbar}
(J_x\beta)^3\tilde L^\prime_{\mathit ME}~\\
\label{dimen1}L^\prime_{\mathit EM}\ &=&\
\frac{J_x\mu_B}{\hbar}\frac{\mu_B c}{\hbar}
 (J_x\beta)^2\tilde L^\prime_{\mathit EM}~,
\end{eqnarray}

where $\tilde{\phantom a}$ denotes
the dimensionless theory-result, $c$ is the distance between
neighbor sites, and $\mu_B$ is the Bohr magneton. 
Note that $L_{\mathit ME}$ has a factor $1/T$ with respect to
$L_{\mathit EM}$; hence a third factor $J_x\beta$ 
in (\ref{dimen}). Inspection of that formula shows
the units of $L_{ME}$ to be ${\rm J(Tesla)^{-2}m/(Ks^{2})}$.

For the sake of completeness we state the analogous results
$$
L_{\mathit MM}\ =\ \frac{J_x\mu_B}{\hbar }\frac{\mu_B c}{\hbar}
(J_x\beta)\tilde L_{\mathit MM}
$$
for the magnetic Drude weight and 
$$ 
L_{\mathit EE}\ =\ \frac{k_BJ_x}{\hbar }\frac{J_x c}{\hbar}
(J_x\beta)^2\tilde L_{\mathit EE}
$$
for the thermal Drude weight. The units can be read off as
$[L_{\mathit MM}]={\rm J (Tesla)^{-2}m/ s^2}$ 
and $[L_{\mathit EE}]={\rm Jm/(Ks^2)}$. Note
that $[\hat L_{ij}]=[L_{ij}]{\rm s}$, see (\ref{L_tau}).

We take the well studied one-dimensional Heisenberg
chain $\rm Sr_2CuO_3$ as an example, \cite{Sol01} 
with the following parameters $c=3.91\,{\rm \AA}$,\cite{busch}
 $J_x/k_B=2.2\cdot10^{3}\,{\rm K}$.
\cite{JSrCuO}. There are
two chains in an area of $12.68\times 3.48{\rm \AA^2}$.\cite{busch}
This leads to a magnon velocity
$v_s=J_x c/\hbar\simeq1.2\cdot10^5\,{\rm m/s}$.
The mean-free path $\lambda_s(T)$ extracted from 
a quasi-classical interpretation of thermal conductivity data \cite{Sol01}
is strongly temperature dependent, it ranges from
$\lambda_s(200{\rm K})\approx 100\,{\rm \AA}$ to
$\lambda_s(50{\rm K})\approx 800\,{\rm \AA}$. This
results via $\lambda_s=v_s\tau$ in a relaxation
time $\tau(200{\rm K})\simeq 0.8\cdot10^{-13}{\rm s}$ 
and
$\tau(50{\rm K})\simeq 6.6\cdot10^{-13}{\rm s}$, in order
of magnitude.

\subsection{Thermopower}
For the {\em thermopower}, we obtain 
$$
Q_M^\prime\ =\ 
{L^\prime_{\mathit ME}\over L_{\mathit MM}} \ =\ 
\frac{k_B}{J_x}(J_x\beta)^2
{\tilde L^\prime_{\mathit ME} \over \tilde L_{\mathit MM}}\ \equiv\
\frac{k_B}{J_x}\tilde Q_M^\prime~.
$$
Typically, $k_B/J_x\approx 10^{-3}{\rm K^{-1}}$ (e.g., in $\rm Sr_2CuO_3$)
and at room temperature $\tilde Q_M^\prime\approx
0.1$. This yields $Q_M^\prime\approx 10^{-4}{\rm K^{-1}}$.

\subsection{Seebeck-effect}
The dimensionless $B$-derivative of the
magnetothermal response is 
$(J_x\beta)^3\tilde L_{\mathit ME}^\prime\simeq0.1$, 
in order of magnitude, see Fig.\ \ref{figlem}.
We therefore obtain from Eq.\ (\ref{dimen})
at $T=200\,{\rm K}$
for the magnetic particle current density a value of 
$$0.036\times B\times\nabla T
\qquad\quad {\rm moments/s}
$$
per Tesla per (Kelvin/m), i.e.,  in SI-units.
For a sample of length $1\,{\rm cm}=10^{-2}\,{\rm m}$
with a temperature difference of $10\,{\rm K}$ and
a field of $10\,{\rm Tesla}$ we
have a current of $360$ spin-1 moments (magnons)
per second. A sample of $\rm Sr_2CuO_3$
with a cross-section of $1\,{\rm mm}^2$ contains
$4.5\cdot10^{12}$ chains. It would induce
a magnetic current of $1.6\cdot10^{15}$ moments per
second. 

\subsection{Closed system}
Finally, we consider again the  setting of a {\em closed system}.
We are given a sample with open ends such that no current may flow. 
A temperature gradient should
therefore lead to a gradient in the magnetization.
Using the Einstein relation for the magnetization current
and Eq.\ (\ref{ME_chi}) we have
$$\nabla M\ =\ \frac{L_{\mathit ME}}{D_{\mathit MM}}\nabla T
\ =\ -B\frac{\partial \chi}{\partial T} \nabla T~.
$$
$\chi B=M$ is the magnetization caused by the presence of a  magnetic field.
Hence, dividing by $M$
$$\nabla M/M\ =\ -\frac{\partial}{\partial T}[\ln\tilde\chi]\nabla T
$$ 
or multiplying both sides with the system length
$$\Delta M/M\ =\ 
-\frac{k_B}{J_x}\frac{\partial}{\partial \tilde T}[\ln\tilde\chi]\Delta T~.
$$ 
Note, that $\Delta M/M$ is not directly a function of 
magnetothermal coefficients.
The magnitude of the dimensionless quantity 
$\frac{\partial}{\partial \tilde T}[\ln\tilde\chi]$ varies about $0.1$.
We therefore find with $J_x/k_B=2200\,{\rm K}$
that the relative change in magnetization
-- from one end of the sample to the other --
should be a fraction of the order of $5\cdot10^{-5}{\rm K^{-1}}\times\Delta T $. 

\subsection{Peltier-effect}
We want to review now briefly the adjoint {\em (Peltier) effect},
described by the
coefficient $L_{\mathit EM}$.
We now turn once again to our reference system $\rm Sr_2CuO_3$.
Here it is instructive to compare the energy currents
driven by a temperature gradient and a gradient in the magnetic field.

We set all quantities as above, with the exception
of the values $(J_x\beta)^2\tilde L_{\mathit EM}^\prime\simeq0.01$,
and  $(J_x\beta)^2\tilde L_{\mathit EE}^\prime\simeq0.1$, which we find
more appropriate. We also do  not compute the energy current density
for one chain but per volume in SI-units.
The result for the thermal response is
$$j_E\ =\ 16\times\nabla T \qquad{\rm\; J/(s\; m^2)}$$
and for the magnetic response
$$j_E\ =\ 3.3\cdot 10^{-4}\times\nabla B\times B \qquad{\rm\; J/(s\; m^2)}.
$$
%

\section{Conclusion} \label{sec_discussion}

We have shown that non-trivial magnetothermal
currents may be induced in one-dimensional
quantum-spin chains in an external magnetic
field. We have argued that these
effects might be especially important for
the 1D Heisenberg chain where the energy-current
operator commutes with the Hamiltonian and the
magnetothermal response diverges. We have presented
estimates of the size for the current of magnetic
moments induced by a thermal temperature difference
for $\rm Sr_2CuO_3$. We believe the size
of the induced magnetic current to be sizeable and
note that the observability of magnetic currents
in magnetic insulators
has been discussed recently \cite{Meier02}.

The support of the German Science Foundation
is acknowledged.

\section{Appendix} \label{sec_appendix}

\subsection{Proof of Eq.\ (\ref{exact_ein})}
\subsubsection{Proof, first step}
We present the proof of Eq.\ (\ref{exact_ein}) in two steps.
First we extend the Hamiltonian to comprise the static response
to a thermal current:
$$H(\lambda,B)=H(B)+\lambda J_E.$$
We indicate the expectation value at nonzero $\lambda$
and $B$ by indices. Moreover, we mean that
these variables are zero, if we omit them
in an index or an argument. `$\Delta$' means subtraction of the
expectation value.

We now show that
\begin{equation}
\langle J_M \Delta H\rangle_{\lambda,B}\ =\
\langle J_E \Delta M\rangle_{\lambda,B}.
\label{proof_one}
\end{equation}
We explicitly assume PBC -- no polarization operators--
 and invoke the equation of continuity
to obtain a local version of the above formula
\begin{eqnarray*}\langle (J_M^{n}-J_M^{n-1})
\Delta h_m\rangle_{\lambda,B}
&=&-\langle i[H,\Delta S_n^z]\Delta h_m\rangle_{\lambda,B}\\
=\langle\Delta  S_n^zi[H,\Delta h_m]\rangle_{\lambda,B}
&=&-\langle \Delta S_n^z  (J_E^{m}-J_E^{m-1})\rangle_{\lambda,B}.
\end{eqnarray*}
On the right-hand side of the above equation
we perform a reflexion in space
$$\langle (J_M^{n}-J_M^{n-1} )\Delta h_m\rangle_{\lambda,B}=
\langle\Delta  S_{-n}^z  (J_E^{-m}-J_E^{-m+1})\rangle_{-\lambda,B}\;. $$
Translational invariance leads for
$$a_{n,m} := \langle J_M^n\Delta h_m\rangle_{\lambda,B} 
\qquad b_{n,m}:=-\langle J_E^n\Delta S_m^z\rangle_{-\lambda,B}
$$
to $a_{n,m}\equiv a_{n-m}$ and to
$$
 a_{k}-a_{k-1}=b_{k+1}-b_{k}\quad\Leftrightarrow\quad
a_{k}-b_{k+1}=a_{k-1}-b_{k},
$$
where $k=n-m$. This last equation implies that
$$
c\ :=\ a_{k}-b_{k+1}
$$
does not depend on the index $k$.
Our strategy is to show that $c$ -- a function on $\lambda $ and
$B$ -- may be neglected with impunity in the thermodynamic limit.
To this end we look at the following estimate:
$$
|c(B,\lambda)|\ =\ |b_{k+1}-a_{k}|
\ \leq \ \min_k\{|b_{k+1}|+|a_{k}|\}.
$$
The fact that all correlation function 
of the form $\langle\Delta G_1\Delta G_2\rangle$ 
(as, e.g., $a_k$ and $b_k$) decay to zero
when the spatial extension (i.e., the index $k$) 
goes to infinity leads to $\lim_{{\rm Vol}\to\infty} c=0$.
Hence we find that in the thermodynamic limit 
$a_{k}=b_{k+1}\Rightarrow \sum a_k=\sum b_k$.
And therefore:
$$\langle J_M  H\rangle_{\lambda,B}/{\rm Vol}\ =\
-\langle J_E  M\rangle_{-\lambda,B}/{\rm Vol}.$$
The initial claim (\ref{proof_one}) follows -- once again --
by a reflexion in space on the second term.

\subsubsection{Proof, second step}
 
Having attained our first goal the further proceeding
is standard.
We apply consecutively derivatives with respect to $\lambda$
and $B$ on our result and obtain: 
%
\[
\langle J_M J_E\Delta H\Delta  M\rangle\ =\
-T\frac{d}{dB}\langle J_E J_E\Delta M\rangle_{B}.
\]
We note that the $\Delta H$ (resp., $\Delta M$) might come
 from a derivative
with respect to $-d/d\beta=T^2d/dT\;$ (resp., $-Td/dB$)
 and use $\langle T_{EM}(-\beta \Delta M)\rangle=-\beta \langle JJ_E(-\beta
\Delta  M)\rangle$
and $\langle -T_{EE}\rangle=\beta^2\langle J_EJ_E\rangle$.
Our formula may then be rewritten as:
\[
\frac{d}{dT}[T^2\langle T_{EM}(-\beta M)\rangle]\ =\
T^2\frac{d^2}{dB^2}\langle -T_{EE}\rangle_{B},
\]
\begin{flushright}Q.E.D.\end{flushright}



\begin{thebibliography}{aa}

\bibitem{hub68} D.L.~Huber,
                Prog. Theo. Phys. {\bf 39}, 1170 (1968);
                D.L.~Huber and J.S.~Semura,
                Phys. Rev. {\bf 182}, 602 (1969).

\bibitem{NieVia71}   Th.~Niemeijer and
                    H.~A.~W.~ van Vianen, Phys. Lett. A {\bf 34}, 401 (1971).

\bibitem{ZNP} X.~Zotos, F.~Naef, and P.~Prelov\v{s}ek, 
              {\it ``Transport and conservation laws''}
              Phys.\ Rev.\ B {\bf 55}, 11029 (1997).

\bibitem{AlGr_kappa} J.V.~Alvarez and C.~Gros,
                  {\it ``Anomalous thermal conductivity of frustrated
                   Heisenberg spin chains and ladders''}, \newline
                   Phys. Rev. Lett. {\bf 89}, 156603 (2002).

\bibitem{hei02} F.~Heidrich-Meisner, A.~Honecker, D.C.~Cabra
                and W.~Brenig,
{\it ``Thermal conductivity of anisotropic and 
frustrated spin-1/2-chains''},
               Phys. Rev. B {\bf 66} 140406 (2002).

\bibitem{KlSa} A.~Kluemper and K.~Sakai,
    {\it ``The thermal conductivity of the spin-1/2 XXZ 
           chain at arbitrary temperature''},
        J.Phys. {\bf A35}, 2173 (2002).


\bibitem{sol00} A.V.~Sologubenko, K.~Giann\`o, H.R.~Ott, U.~Ammerahl and
                A.~Revcolevschi,
                Phys. Rev. Lett. {\bf 84}, 2714 (2000).

\bibitem{hes01}  C.~Hess, C.~Baumann, U.~Ammerahl, B.~B\"uchner,
                 F.~Heidrich-Meisner, W.~Brenig and A.~Revcolevschi,
                 Phys. Rev. B {\bf 64},  184305 (2001).

\bibitem{Sol01} A.V.~Sologubenko, K.~Giann\`{o}, H.R.~Ott, 
                A.~Vietkine and A.~Revcolevschi,
 {\it ``Heat transport by lattice and spin excitations in the spin-chain
         compounds SrCuO$_2$ and Sr$_2$CuO$_3$''}, 
          Phys. Rev. B {\bf 64}, 054412 (2001)

\bibitem{thu01} K.R.~Thurber, A.W.~Hunt, T.~Imai and F.C.~Chou,
{\it `` $^{17}$O NMR Study of q=0 Spin Excitations in a Nearly Ideal S=1/2
    1D Heisenberg Antiferromagnet, Sr2CuO3, up to 800 K''},
    Phys. Rev. Lett. {\bf 87}, 247202 (2001).

\bibitem{ChLu} P.~M.~Chaikin and T.~C.~Lubensky, 
              {\it``Principles of condensed matter physics''},
              Cambridge University Press, Cambridge, England 1995.




\bibitem{Abrikosov} A.~A.~Abrikosov, 
                    {\it ``Fundamentals of the theory of metals''}, 
                    North-Holland 1987; H.~Ashcroft and N.~Mermin,  
                   {\it ``Solid state physics''}, 
	           Holt-Saunders (1976).

\bibitem{ZMR} D.~Zubarev, V.~Morozov, and G.~Roepke,  
              {\it ``Statistical mechanics of nonequilibrium processes''},
              Akademie Verlag, Berlin 1997.


\bibitem{AlGr_MC} J.V.~Alvarez and C.~Gros ,
                 {\it ``Conductivity of quantum-spin chains: A
                 Quantum Monte Carlo approach''},
		 Phys. Rev. B {\bf 66} 094403 (2002).

\bibitem{Shastry} B.~S.~Shastry and B.~Sutherland,
       {\it ``Twisted boundary conditions and effective mass in
              Heisenberg-Ising and Hubbard rings''},
             Phys. Rev. Lett. {\bf 65}, 243 (1990).

\bibitem{Barnard} R.~D.~Barnard, 
                  {\it``Thermoelectricity in metals and alloys''},
                  Taylor \& Francis ltd, 1972.

\bibitem{SWZ} D.~J.~Scalapino, S.~R.~White and S.~Zhang, 
              {\it ``Insulator, metal, or superconductor: The criteria''},
              Phys.\ Rev.\ B {\bf 47}, 7995 (1993).

\bibitem{GRMA} M.~P.~Grabowski and P.~Mathieu, Ann. Phys. {\bf 243}, 299, (1995).

\bibitem{HalShas} R.B.Laughlin {\it et al.}, in ``Field Theories for 
           Low-Dimensional Condensed Matter Systems'' edited by
        G.~Morandi {\it et al.},
           Springer Series in Solid-State Sciences 131.

\bibitem{San_loop} A.W.~Sandvik,
                {\it ``Stochastic series expansion method with
                operator-loop update''},
                Phys. Rev. B {\bf 59}, R14157 (1999).

\bibitem{Kl_Drude} A.~Kl\"umper, private communications and
                   J.~Benz, T.~Fukui, A.~Kl\"umper, and C.~Scheeren
		   (in preparation).



\bibitem{zot99} X.~Zotos, 
      {\it ``Finite Temperature Drude Weight of the
      One-Dimensional Spin-1/2 Heisenberg Model''},\newline
      Phys. Rev. Lett. {\bf 82}, 1764  (1999).

\bibitem{AlGr_Drude} J.V.~Alvarez and C.~Gros,
            {\it ``Low-temperature transport in Heisenberg chains''},
            Phys. Rev. Lett. {\bf 88}, 077203 (2002).

\bibitem{NMA} B.~N.~Narozhny, A.~J.~Millis, and N.~Andrei, Phys. Rev. B
	{\bf 58}, R2921 (1998).



\bibitem{busch}  C.~L.~Teske and H.~M\"uller-Buschbaum,
	Z. Anorg. Allg. Chem. {\bf 371}, 325 (1969).


\bibitem{JSrCuO} T.~Ami, M.~K.~Crawford, R.~L.~Harlow, Z.~R.~Wang,
        D.~C.~Johnston, Q.~Huang, and R.~W.~Erwin,   
       Phys. Rev. B {\bf 51}, 5994 (1995);
       H.~Suzuura, H.~Yasuhara, A.~Furusaki, N.~Nagaosa, and
       Y.~Tokura, Phys. Rev. Lett. {\bf 76}, 2579 (1996);
       N.~Motoyama, H.~Eisaki, and S.~Uchida,  
      Phys. Rev. Lett. {\bf 76}, 3212 (1996);
       D.~C.~Johnston, Acta Phys. Pol A {\bf 91}, 181 (1997).


\bibitem{Meier02} F. Meier and D. Loss,
{\it ``Magnetization transport and quantized spin conductance''},
cond-mat/0209521.


\end{thebibliography}
\end{document}